\documentstyle[psfig]{spackap}
\begin{document}
\bibliographystyle{klunamed}
\newcommand {\bpic} {{$\beta\,$Pic~}}
\newcommand {\bpi} {{$\beta\,$Pic}}
\newcommand {\alyr} {{$\alpha\,$Lyr~}}
\newcommand {\aly} {{$\alpha\,$Lyr}}
\newcommand {\apsa} {{$\alpha\,$PsA~}}
\newcommand {\eeri} {{$\epsilon\,$Eri~}}
\newcommand {\lboo} {{$\lambda\,$Boo~}}
\newcommand {\um} {$\mu$m\ }
\newcommand {\ums} {$\mu$m}
\newcommand {\umm} {\mu{\rm m}\,}
\newcommand {\etal} {{\em et.\ al.} }
\newcommand {\apgt} {\ {\raise-.5ex\hbox{$\buildrel>\over\sim$}}\ }
\newcommand {\aplt} {\ {\raise-.5ex\hbox{$\buildrel<\over\sim$}}\ }
\newcommand {\degr} {$^\circ$\,}
\begin{article}
\begin{opening}
\title{Galactic and Extragalactic Magnetic Fields}
\author{RAINER \surname{BECK}\vspace{0.15cm}}
\institute{Max-Planck-Institut f\"ur Radioastronomie, Auf dem
H\"ugel 69, D-53121 Bonn, Germany}
\begin{ao}
Max-Planck-Institut f\"ur Radioastronomie, Auf dem H\"ugel 69,
D-53121 Bonn, Germany; {\tt rbeck@mpifr-bonn.mpg.de}
\end{ao}
\runningauthor{RAINER BECK}
\runningtitle{GALACTIC AND EXTRAGALACTIC MAGNETIC FIELDS\vspace{-0.3cm}}
\date{Received: 14 July 2000; Accepted: 15 November 2000\vspace{-0.4cm}}
\begin{abstract}
The current state of research of the Galactic magnetic field is reviewed
critically. The average strength of the {\it total} field derived from radio
synchrotron data, under the energy equipartition assumption, is $6\pm2\mu$G
locally
and about $10\pm3\mu$G at 3~kpc Galactic radius. These values agree well with
the estimates using the locally measured cosmic-ray energy spectrum and
the radial variation of protons derived from $\gamma$-rays. Optical and
synchrotron
polarization data yield a strength of the local {\it regular} field of
$4\pm1\mu$G,
but this value is an upper limit if the field strength fluctuates within the
beam or if anisotropic fields are present. Pulsar rotation measures,
on the other hand, give only $1.4\pm0.2\mu$G, a lower limit if
fluctuations in regular field strength and
thermal electron density are anticorrelated along the pathlength.

The local regular field may be part of
a ``magnetic arm'' between the optical arms. However,
the global structure of the regular Galactic field is not yet known.
Several large-scale field reversals in the Galaxy were detected from
rotation measure data, but a similar phenomenon was not observed in
external galaxies. The Galactic field may be young in terms of dynamo
action so that reversals from the chaotic seed field are preserved,
or a mixture of dynamo modes causes the reversals,
or the reversals are signatures of large-scale anisotropic field loops.
The Galaxy is surrounded
by a thick disk of radio continuum emission of similar extent
as in edge-on spiral galaxies. While the local field in the thin disk
is of even symmetry with respect to the plane (quadrupole), the
global thick-disk field may be of dipole type.
The Galactic center region hosts highly regular fields of up to milligauss
strength which are oriented perpendicular to the plane.

A major extension of the data base of pulsar rotation measures
and Zeeman splitting measurements is
required to determine the structure of the Galactic field.
Further polarization surveys of the Galactic plane at wavelengths of
6~cm or shorter may directly reveal the fine structure of the local
magnetic field.
\end{abstract}
\end{opening}
\setcounter{page}{1}
\volume{...}
\newcommand{\be}{\begin{equation}}
\newcommand{\ee}{\end{equation}}

\section{Observational Methods}

Interstellar magnetic fields can be observed indirectly at
optical, infrared, submillimeter and radio wavelengths. Extensive
reviews of observational methods were given by
\inlinecite{Heiles1976}, \inlinecite{Spoelstra1977},
\inlinecite{Verschuur1979} and \inlinecite{Tinbergen1996}. Zeeman
spectral-line splitting data are available for gas clouds, e.g.
near the Galactic center (Section~6). Optical polarization data
yield the large-scale structure of the magnetic field in the local
spiral arm (Section~3.1), but small-scale features may be
contaminated by scattered light which is unrelated to magnetic
fields. Polarization observations in the infrared and
submillimeter ranges are rapidly evolving (see e.g.
\opencite{Greavesetal2000}) and are expected to contribute to our
knowledge about magnetic fields in the near future. Linearly
polarized radio continuum emission at centimeter wavelengths and
its Faraday rotation provide the most extensive and reliable
information on large-scale interstellar magnetic fields in
external galaxies (\opencite{Becketal1996}; \opencite{Beck2000})
and in the Galactic center (Section~6). Information about the
large-scale structure of the Galactic magnetic field near the Sun
mainly comes from Faraday rotation measures of pulsars and
extragalactic sources (Section~3) while polarization surveys of
the Galactic plane at centimeter wavelengths reveal a wealth of
small-scale magnetic structures.

Single-dish telescopes, like the 100-m Effelsberg dish, are powerful
instruments
in detecting weak, extended radio emission. To achieve sufficiently high
resolution in external galaxies, interferometric (synthesis) telescopes
are used (like the Westerbork Synthesis Radio Telescope, the Very Large
Array,
the Australia Telescope Compact Array) which, however, miss large-scale
structures so that the combination of interferometric with single-dish
data is often required.

Interstellar magnetic fields are illuminated by cosmic-ray
electrons, spiralling around the field lines and emitting
synchrotron radiation, the dominant contribution to the diffuse
radio continuum emission at centimeter and decimeter wavelengths.
Synchrotron emission is intrinsically highly linearly polarized,
70--75\% in a completely regular magnetic field. The observable
degree of polarization in galaxies is reduced by Faraday
(wavelength-dependent)  depolarization in magnetized plasma
clouds, by geometrical \linebreak (wavelength-independent) depolarization due
to variations of the magnetic field orientation across the
telescope beam and along the line of sight, and by a contribution
of unpolarized thermal emission which is dominating only in
star-forming regions of the Galaxy. A map of the total radio
intensity reveals the {\it strength of the total interstellar
magnetic field in the plane of the sky} (averaged over the volume
traced by the telescope beam), polarized intensity and
polarization angle (see Figure~4) the {\it strength and structure
of the resolved regular field in the plane of the sky}.

Polarization angles are ambiguous by $\pm 180$\degr and hence
insensitive to field reversals. Imagine that a magnetic field
without any regular structure (an isotropic random field) is
compressed or stretched in one dimension. Emission from the
resulting anisotropic field is linearly polarized with ordered
vectors, but the field is {\it incoherent}, i.e. it reverses its
direction frequently within the telescope beam. Here the
polarization vectors just indicate {\it anisotropy} of the
magnetic field distribution in the emission region.

The orientation of polarization vectors is changed in a
magneto-ionic medium by {\it Faraday rotation} which is
proportional to the product of the average density of thermal
electrons and the strength of the regular field component along
the line of sight. At centimeter wavelengths the Faraday rotation
angle ($\Delta \chi$) of the polarization vectors varies with
$\lambda^2$. ($ \Delta \chi = RM \, \lambda^2$, where $RM$ is
called the {\it rotation measure}, see Section~2.2.) Typical
interstellar rotation measures of $\rm\simeq 50\, rad/m^2$ lead to
$126$\degr rotation at $\lambda21$~cm, $10$\degr at $\lambda6$~cm
and $3$\degr at $\lambda3$~cm. Below about $\lambda3$~cm Faraday
rotation is generally small so that the $\mathbf{B}$--vectors
(i.e. the observed $\mathbf{E}$--vectors rotated by $90$\degr)
directly trace the {\it orientation\/} of the regular field in the
sky plane. However, Faraday rotation near the Galactic center is
significant even at $\lambda3$~cm (Section~6). Maps of Faraday
rotation measures (see Figures~3 and 5) give the strength and the
{\it direction} of the average field components along the line of
sight. $RMs$ are essential to distinguish between coherent and
incoherent fields: In an incoherent (anisotropic) field the $RMs$
are random and show no large-scale structure (see Section~3.2).

At decimeter wavelengths {\it Faraday depolarization}
significantly affects the polarized radio emission
(\opencite{Sokoloffetal1998}). {\it Differential Faraday rotation}
occurs within the synchrotron-emitting medium along the line of
sight even if the magnetic field is completely regular. It leads
to zero polarized intensity (``Faraday shadows'') at certain
wavelengths where the observed rotation of the polarization angle
reaches multiples of $90$\degr. Many ``filaments'' of vanishing
polarized intensity, accompanied by $90$\degr jumps in
polarization angle across the filament, were detected in the
$\lambda21$~cm polarization survey out of the Galactic plane
(\opencite{Uyanikeretal1999}), in high-resolution maps of the
local Galactic emission at $\lambda90$~cm
(\opencite{Haverkornetal2000}) and in polarization maps of
external galaxies (\opencite{Beck2000}). These do {\it not}
indicate regions with zero strength of the regular field, but
large changes in Faraday rotation measures across the
``filament''. Turbulent magnetic fields cause a {\it dispersion}
in Faraday rotation measures which further depolarizes the
emission. Even polarization maps of the Galactic plane at
$\lambda11$~cm (\opencite{Duncanetal1999}) are affected by Faraday
depolarization. The fine structure of the Galactic magnetic field
may appear at wavelengths of $\le\lambda6$~cm, but such
observations are difficult due to the low synchrotron intensities
at short wavelengths.

\section{Magnetic Field Strength in the Galactic Disk}
\subsection{Equipartition}

The average strength of the total $\langle
B_{\mathrm t,\perp}\rangle$ and the resolved regular field
$\langle B_{\mathrm reg,\perp}\rangle$ in the plane of the sky
can be derived from the total and polarized radio synchrotron
intensity, respectively, if energy-density equipartition between
cosmic rays and magnetic fields or minimum total energy density is
assumed. Furthermore, the ratio $K$ between cosmic-ray protons and
electrons (and its variation with particle energy), the
synchrotron spectral index $\alpha$
(i.e. flux density $S\propto\nu^{-\alpha}$), the extent of the
radio-emitting region along the line of sight, and the volume
filling factor of the field have to be known. Fortunately, the
derived equipartition field strength depends on the power
$1/(\alpha +3)$ of each of these parameters so that even large
uncertainties lead to only moderate errors in field strength.

The field strength may vary within the observed volume, e.g. due
to compression by density waves, supernova remnants or flux
freezing in gas clouds. In such cases the equipartition field
strength is {\it overestimated} because $\langle
B_{\mathrm t,\perp}^{3+\alpha}\rangle^{1/(3+\alpha)}$ is larger
than the true mean field $\langle B_{\mathrm t,\perp}\rangle$.
For strong fluctuations ($\delta B/<B> \, \simeq 1$) the
equipartition strength is $\simeq 40\%$ too large. If, on the
other hand, the field is concentrated in filaments with a volume
filling factor $f$, the equipartition/minimum-energy estimate is
smaller than the true field strength in the filaments by a factor
$f^{1/(3+\alpha)}$.

The textbook equipartition/minimum-energy formulae use a fixed
integration interval in radio {\it frequency} to determine the
total energy density of cosmic-ray electrons. This procedure makes
it difficult to compare field strengths between galaxies because a
fixed frequency interval corresponds to different electron energy
intervals if the field strengths are not the same. When instead a
fixed integration interval in {\it energy} is used (say, from
300~MeV to infinity), the minimum-energy and energy equipartition
estimates give very similar values for $\langle
B_{\mathrm t,\perp}^{3+\alpha}\rangle$, where $\alpha$ is
typically $\simeq 0.9$.

The mean equipartition strength of the total field (using $K=100$)
for a sample of 74 spiral galaxies (corrected for the inclination
of each galaxy) is $\langle B_{\mathrm t}\rangle = 9\,\mu$G with a
standard deviation of $3\,\mu$G (\opencite{Niklas1995}). The
values for the Galaxy (Figure~1) fit well into this range. In
nearby galaxies the average equipartition strength of the total
field in the galactic disks ranges between $\langle
B_{\mathrm t}\rangle \simeq 6\,\mu$G in radio-faint spiral galaxies
like M31 and M33, and $\simeq 15\,\mu$G in grand-design galaxies
like M51, M83 and NGC~6946. In spiral arms the total field
strength is $10-20\,\mu$G over kpc scales, but locally can reach
$\simeq1\,$mG in molecular clouds compressed by supernova
shocks (\opencite{Broganetal2000}) and in filaments near the Galactic
center (see Section~6).

Figure~1 shows the radial variation of the equipartition field
strength in the Galaxy, obtained from the radial variation of the
total synchrotron emission at 408~MHz ($\lambda74$~cm) derived by
\inlinecite{Beuermannetal1985} from the survey of
\inlinecite{Haslametal1974}. The total field strength is $6\pm2\,\mu$G
locally and $10\pm3\,\mu$G at 3~kpc Galactic radius.

\begin{figure}
\centerline{\psfig{file=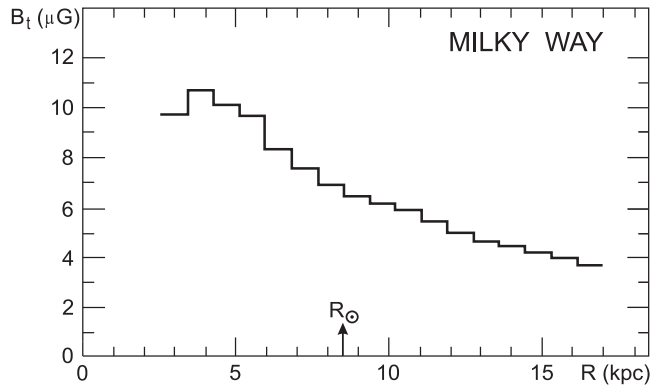,width=8.5truecm,clip=}}
\vspace{0.2cm}
\caption[]{ Strength of the total magnetic field in the Galaxy,
averaged from the deconvolved surface brightness of the
synchrotron emission at 408~MHz (\opencite{Beuermannetal1985}),
assuming energy equipartition between magnetic field and cosmic
ray energy densities (Berkhuijsen, personal communication). The
accuracy is about 30\%. The Sun is assumed to be located at
R=8.5~kpc.}
\end{figure}

The equipartition/minimum-energy assumption is under continuous debate.\\
Firstly, this assumption may be valid only on long
time/space scales where equilibrium conditions can develop,
but may break down on short/small scales.
Secondly, synchrotron or inverse-Compton energy losses of the
cosmic-ray electrons cause a steepening of the GeV electron spectrum
so that $K$ increases with increasing energy. This leads to an
overestimate of the field strength (\opencite{Pohl1993}). However, there
is little indication for a steepening of the radio synchrotron spectrum
at frequencies beyond 1~GHz (and hence the electron spectrum beyond
$\simeq 3$~GeV) of spiral galaxies (\opencite{NiklasandBeck1997}).

In the Galaxy the accuracy of the equipartition assumption can be
tested because we have independent information about the local
cosmic-ray electron energy density from direct measurements and
about the cosmic-ray proton distribution from $\gamma$-ray data.
Combination of the radio synchrotron emission, the local
cosmic-ray electron density and diffuse continuum $\gamma$-rays
yields a local strength of the total field of $6\pm1\, \mu$G
(\opencite{Strongetal2000}), almost the same value as derived from
energy equipartition (Figure~1). Even the radial variation of the
equipartition field strength in Figure~1 is in excellent agreement
with that derived by \inlinecite{Strongetal2000}.

Synchrotron polarization observations in the Galaxy at decimeter
wavelengths imply a ratio of regular to total field strengths
within about a kpc from the Sun of $ \, <
B_{\mathrm reg}/B_{\mathrm t}
> \, \simeq 0.6$ (\opencite{Berkhuijsen1971};
\opencite{BrouwandSpoelstra1976}; \opencite{Heiles1996}). The
radio emission along the local spiral arm requires that the
strength of the regular field components is similar or slightly
less than that of the turbulent field, or $ \, <
B_{\mathrm reg}/B_{\mathrm t}
> \,\, \le 0.7$ (\opencite{Phillippsetal1981}). For $\langle
B_{\mathrm t}\rangle = 6\pm2\,\mu$G these results give $4\pm1\,\mu$G
for the large-scale regular component.

Experience from external galaxies shows that the regular field is
weaker and the turbulent field is stronger in the spiral arms, probably
due to field tangling by star-forming processes and the expansion of
supernova remnants. In interarm regions the regular field can be
much stronger than the turbulent field.

The strength of the large-scale regular fields $B_{\mathrm reg}$ in
spiral galaxies (observed with a spatial resolution of a few
100~pc) is typically 1--5$\,\mu$G, up to $\simeq 13\,\mu$G in an
interarm region of NGC~6946 hosting an exceptionally strong
regular field (\opencite{BeckandHoernes1996}) (see Figure~4).

Note that polarized emission does not distinguish between coherent
and inhoherent (anisotropic) fields (Section~1) so that the
strength of the coherent regular field may be lower than the
equipartition estimates.

\subsection{Pulsars}

The Faraday rotation measure is defined as $ RM \, = \, k  \, \int
\, n_{\mathrm e}  \, \, B_{\|}  \, \, dl  $, where $B_{\|}$ is the
component of the regular field $B_{\mathrm reg}$ along the line of
sight. Measurement of pulsar $RMs$ together with their dispersion
measures $DM  =  \int  n_{\mathrm e}   dl$ allows an estimate of
$B_{\|}$ of the local Galactic field (the ``standard estimate''):\\
$ <B_{\|}> \, = \, RM \, DM^{-1} \, k^{-1} \,\, $.

The strength of the local regular field is $ <B_{\mathrm reg}> \,
= 1.4 \pm 0.2\, \mu G$ (\opencite{RandandLyne1994};
\opencite{HanandQiao1994}; \opencite{IndraniandDeshpande1998}),
less than the equipartition estimate (Section~2.1).

The above field estimate suffers from various problems:
\begin {itemize}
\item The ``standard estimate'' assumes that the variations of $B_{\|}$
and $n_{\mathrm e}$ are uncorrelated along the line of sight. If
however these quantities are {\it correlated}, we get
$ RM/DM \, = \, k \,  <B_{\|}> ( 1 \, + \, \delta B_{\|}^2 / <B_{\|}>^2 )
$.\\
If $\,\, \delta B_{\|} / <B_{\|}> \simeq 0.5 \, \,$:
\hspace*{0.5cm} $ <B_{\|}> \, = \, 0.8 \, RM \,DM^{-1} \, k^{-1} $.\\
Here the standard estimate is too large and even increases the discrepancy
to the equipartition value.
\item Observations of external galaxies indicate that $B_{\mathrm reg}$
and
$n_{\mathrm e}$ are {\it anticorrelated} on kpc scales
(\opencite{BeckandHoernes1996}). $B_{\mathrm reg}$ is strongest in
the interarm regions and weaker in the spiral arms where
$n_{\mathrm e}$ is largest (see Figure~4). Even if $<n_{\mathrm e}>$
and $<B_{\mathrm reg}>$ are uncorrelated in different regions of
the sky, the fluctuations $\delta n_{\mathrm e}$ and $\delta B$ can
still be anticorrelated on small scales along the line of sight
(pressure balance), e.g.  $\delta B_{\|}/<B_{\|}> = - \delta
n_{\mathrm e}/<n_{\mathrm e}>$. For small
fluctuations we get:\\
$ RM/DM \, = \, k \,  <B_{\|}> ( 1 \, - \, \delta B_{\|}^2 / <B_{\|}>^2 )
$.\\
If $\,\, \delta B_{\|} /  <B_{\|}> \simeq 0.5 \, \,$:
\hspace*{0.5cm} $ <B_{\|}> \, = \, 1.3 \, RM \,DM^{-1} \, k^{-1} $.\\
Here the standard estimate is too small.
{\it The larger the anticorrelated fluctuations, the smaller the observed
rotation
measure.}
\item If there are $N$ field {\it reversals} along the line of sight,
the standard estimate is $\simeq(N+1)$times too small. The pulsar $RM$ data
have been corrected for detected reversals, but there may be more
{\it hidden reversals}.
\end{itemize}

Hence $<B_{\mathrm reg}>$ derived from pulsar $RMs$ may be an
underestimate. On the other hand, equipartition field strengths
based on polarized synchrotron emission may be too large
(Section~2.1). Anisotropic fields (produced e.g. by shear motions,
elongated supernova remnants, or Parker loops) may explain
the discrepancy in field strengths and also some of the
reversals near the Sun (see Section~3.2).

\section{Structure of the Regular Field in the Galactic Disk}

Highly polarized radio emission from the local spiral arm was observed at
decimeter
wavelengths with the Dwingeloo single-dish telescope (e.g.
\opencite{Berkhuijsen1971};
\opencite{BrouwandSpoelstra1976}). A substantial fraction of the Galactic
magnetic field must be regular on various spatial scales. With higher
spatial resolution more fine structure of the field can be resolved.
Surveys of the Galactic plane with the Parkes telescope
(\opencite{Duncanetal1997}) and the Effelsberg telescope
(\opencite{Junkesetal1987}; \opencite{Duncanetal1999};
\opencite{Uyanikeretal1999}) revealed high degrees of polarization
in small nearby regions where the regular field dominates.
In external galaxies average fractional polarizations, observed with several
100~pc spatial resolution, are less than
a few percent in central regions and spiral arms, but 20--40\% in {\it
between}
the spiral arms and in outer regions where the field is regular on
very large scales (\opencite{Becketal1996}; \opencite{Beck2000}).

\subsection{Pitch Angles}

The Sun is located between two spiral arms, the Sagittarius/Carina
and the Perseus arms, as delineated e.g. by thermal electrons
(\opencite{TaylorandCordes1993}; \opencite{Vallee1996}). The
local (or Orion) arm seen in optical and HI emission is probably
an interarm spur. The mean pitch angle of the local and
neighboring spiral arms is $\simeq -18$\degr for the stars and
$\simeq -13$\degr for the gas (see compilation by
\opencite{Vallee1995}). However, starlight polarization and
pulsar $RM$ data (Figure~2) give a significantly smaller pitch
angle ($\simeq -8$\degr) for the local magnetic field
(\opencite{Heiles1996}; \opencite{HanandQiao1994};
\opencite{IndraniandDeshpande1998}; \opencite{Hanetal1999a}).

The pitch angle of the ``magnetic arms'' in the spiral galaxy NGC
6946 (see Figure~4) is $10$--$20$\degr, smaller than that of the
optical arms (\opencite{Rohdeetal1999}). The local Galactic field
may also form a ``magnetic arm'' with a small pitch angle, located
between two optical arms (\opencite{HanandQiao1996}). The
amplitude of the $RM$ variation with azimuthal angle in NGC~6946
(see Figure~5) is $\rm\simeq 70\, rad/m^2$ at an inclination of
$30$\degr. Looking along the field, we would observe $RM\simeq
140\, {\rm rad/m}^2$, similar to $RMs$ of extragalactic sources
seen along the local arm (Figure~3).

\subsection{Field Reversals}

The most interesting property of the Galactic magnetic field is
the existence of {\it reversals}. A reversal inside the solar
radius, between the local and the Sagittarius arm, was already
detected from rotation measures of polarized extragalactic radio
sources (\opencite{SimardandKronberg1980}). A recent analysis with
the wavelet technique (\opencite{Fricketal2000}) confirmed this
reversal at $\simeq 0.6$~kpc from the Sun and indicates another
reversal in the outer galaxy. Both reversals are also seen in the
data of pulsar $RMs$ (\opencite{RandandLyne1994};
\opencite{HanandQiao1994}). Two more reversals, one further inside
and one further outside, were suggested by
\inlinecite{Hanetal1999a}.

\begin{figure}
\centerline{\psfig{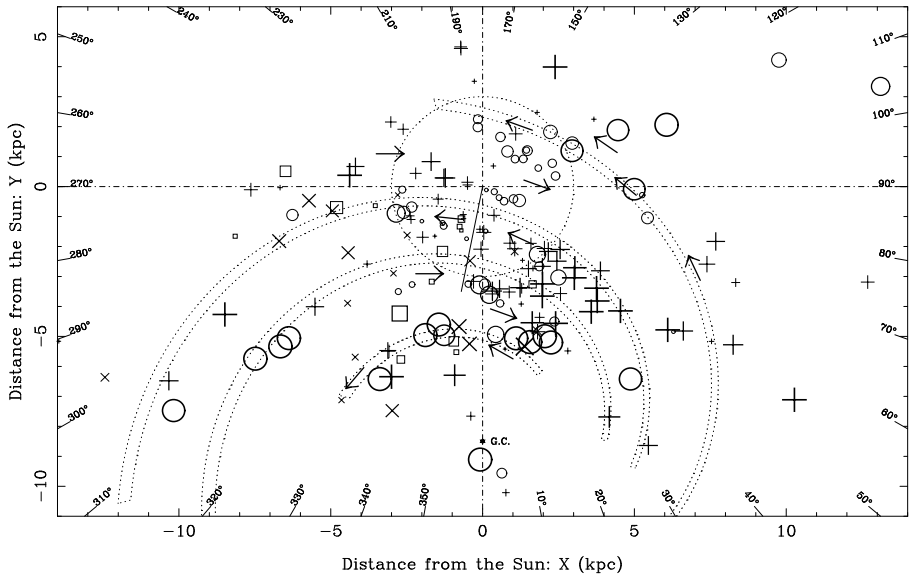}}
\vspace{0.2cm}
\caption[]{The distribution of the $RMs$ of pulsars within
$8$\degr of the Galactic plane. Positive $RMs$ are shown as
crosses, negative $RMs$ as circles. The most recent $RM$ data are
indicated by X and open squares. The symbol sizes are proportional
to the square root of $|RM|$, with the limits of 5 and $\rm
250\,rad/m^2$. The directions of the bisymmetric field model are
given as arrows. The approximate location of four spiral arms is
indicated as dotted lines. The dotted circle has a radius of 3~kpc
(from \opencite{Hanetal1999a}).}
\end{figure}

\begin{figure}
\centerline{\psfig{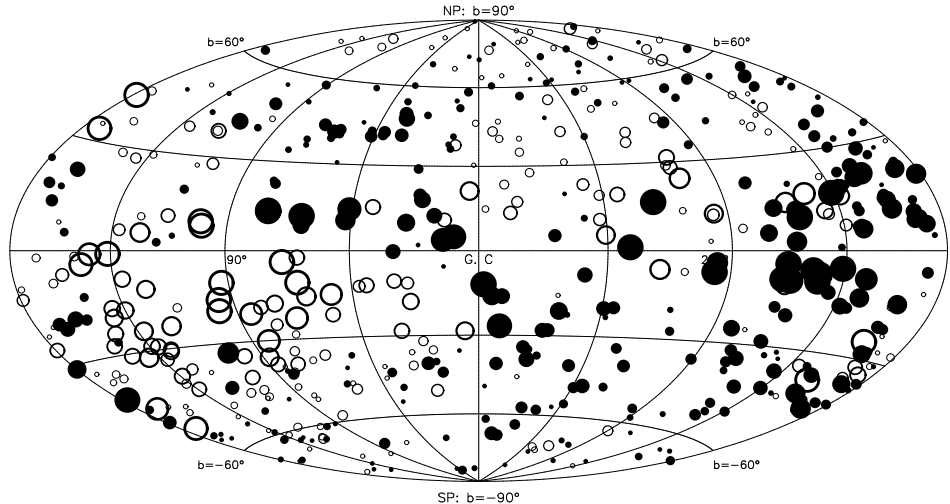}}
\vspace{0.2cm}
\caption[]{The distribution of the $RMs$ of extragalactic radio
sources. Filled circles indicate positive $RMs$, open circles
negative $RMs$. The area of the circles is proportional to $|RM|$
within limits of 5 and $\rm 150\, rad/m^2$ (from
\opencite{Hanetal1997}).}
\end{figure}

Several explanations of the field reversals were proposed. (a)
Dynamos prefer to generate axisymmetric modes without reversals,
but their growth time can be quite long so that elongated field
loops with reversals may remain from the chaotic seed field
(\opencite{Poezdetal1993}). (b) The dynamo may generate higher
modes, e.g. a bisymmetric one with large-scale reversals from one
magnetic arm to the next. To account for several reversals along
Galactic radius, a bisymmetric magnetic spiral with a small pitch
angle ($\simeq -7$\degr) is needed (\opencite{HanandQiao1994};
\opencite{IndraniandDeshpande1998}). Figure~2 shows the
distribution of $RMs$ of low-latitude pulsars, projected onto the
Galactic plane, and the bisymmetric model by
\inlinecite{Hanetal1999a}. Some data disagree with the model. This
may be due to local disturbances of the field e.g. by supernova
remnants (\opencite{Vallee1996}), or the model is too simple.
More pulsar $RM$ data are needed to obtain a better sampling of
the field structure. (c) Nonlinear dynamo models revealed a
mixture of magnetic modes, while the dominance of the bisymmetric
mode is very difficult to obtain. A model based on the rotation
curve of M51 and a spiral modulation generated a large-scale
reversal near the corotation radius in one half of the galaxy
where the bisymmetric field can be trapped by the spiral pattern
over the galaxy's lifetime (\opencite{Bykovetal1997}, see below).
However, no reversals at other radii appeared. (d) Large-scale
anisotropic field loops may be produced by stretching or
compressing (see below).

In external galaxies, data sampling is much denser than in the
Galaxy. High-resolution maps of Faraday rotation, which measure
the $RMs$ of the diffuse polarized synchrotron emission, are
available for a couple of spiral galaxies
(\opencite{Becketal1996}; \opencite{Beck2000}). It is striking
that {\it only very few field reversals} have been detected in
spiral galaxies where the spatial resolution is better than 1~kpc.
The observed disk field of M51 can be described by a mixture of
axisymmetric and bisymmetric components which may mimic a reversal
for an observer located within the disk (see Figure~8a in
\opencite{Berkhuijsenetal1997}, compare with Figure~6 in
\opencite{Bykovetal1997}). In NGC~2997 a reversal between the disk
field and the central region occurs at about 2~kpc radius
(\opencite{Hanetal1999b}), but no reversal like that seen in the
Milky Way was found in the outer disk of any galaxy. Some evidence
for a dominating bisymmetric field structure was found from the
large-scale azimuthal pattern variation of $RMs$ in M33 and M81
(\opencite{Krause1990}), but with recently improved observations
of these galaxies the evidence for M33 has not been confirmed
(\opencite{Fletcheretal2000}).

The discrepancy between Galactic and extragalactic data may be due to
the different volumes traced by the observations. Results in the Galaxy
are based on pulsar $RMs$ which trace only the warm ionized medium near the
plane, while extragalactic results are based on $RMs$ of the diffuse
polarization emission integrated over the whole volume occupied by
regular fields and thermal gas. Reversals in a thin disk would be difficult
to detect in external galaxies -- but why should reversals occur only
near the plane?

\begin{figure}
\centerline{\psfig{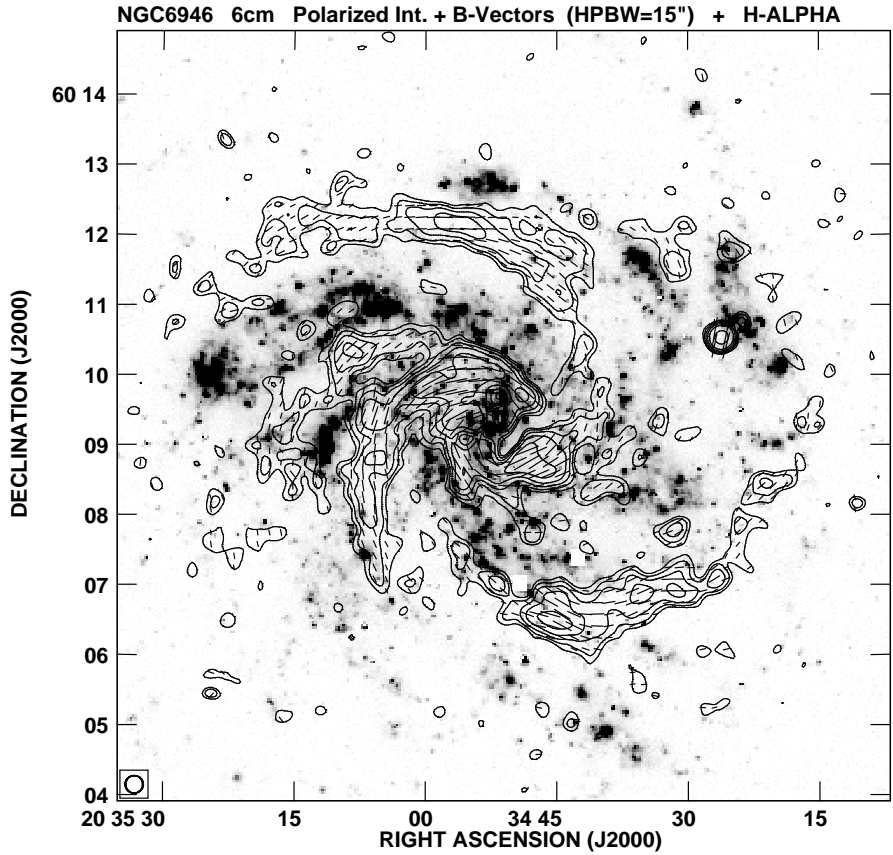}}
\caption[]{Polarized radio emission and $\mathbf{B}$--vectors of
the spiral galaxy NGC~6946 at $\lambda$6~cm, combined from
observations with the VLA and Effelsberg telescopes (from
\inlinecite{Becketal1996} and overlaid onto an H$\alpha$ image of
\inlinecite{Fergusonetal1998}). The angular resolution of the
radio image is 15~arcsec.}
\end{figure}

With the limited number of pulsar $RMs$ it is not sure whether the
reversals in the Galaxy really continue over a substantial part of
the disk. Instead, they may be relics of the protogalactic seed
field or part of a large-scale anisotropic field, as indicated
from observations in external galaxies. Figure~4 shows the
observed distribution of polarized synchrotron emission in the
galaxy NGC~6946 hosting a grand-design spiral field. The
large-scale regular fields are restricted to the interarm region
between the optical arms while the total field in the optical arms
is strong, but tangled. The map of $RMs$ (Figure~5) can only give
data where the signal-to-noise ratio of polarized intensities are
large enough. A large-scale pattern is apparent: $RMs$ are
predominantly positive in the northern half and predominantly
negative in the southern half so that the field points {\it
inwards} (towards the galactic center) everywhere. There is no
large-scale reversal between the outer and the inner ``magnetic
arms'' which are about 3~kpc apart. On smaller scales, $RMs$
fluctuate significantly with a mean rms of 30--50~$\rm rad/ m^2$,
larger than expected from instrumental noise. Furthermore, the
spatial scale of these fluctuations is about 1~kpc, large than
that of the turbulent field and that of instrumental noise which
should have a spatial scale of the telescope resolution which
corresponds to $\simeq 400$~pc in NGC~6946. $RM$ variations in
other spiral galaxies are similar, probably signature of
kpc--scale fluctuations in the field structure (anisotropic fields,
see Section~2.2). For example, a magnetic loop was identified in M31
and interpreted as a Parker loop (\opencite{Becketal1989}).

\begin{figure}
\vspace{0.2cm}
\centerline{\psfig{file=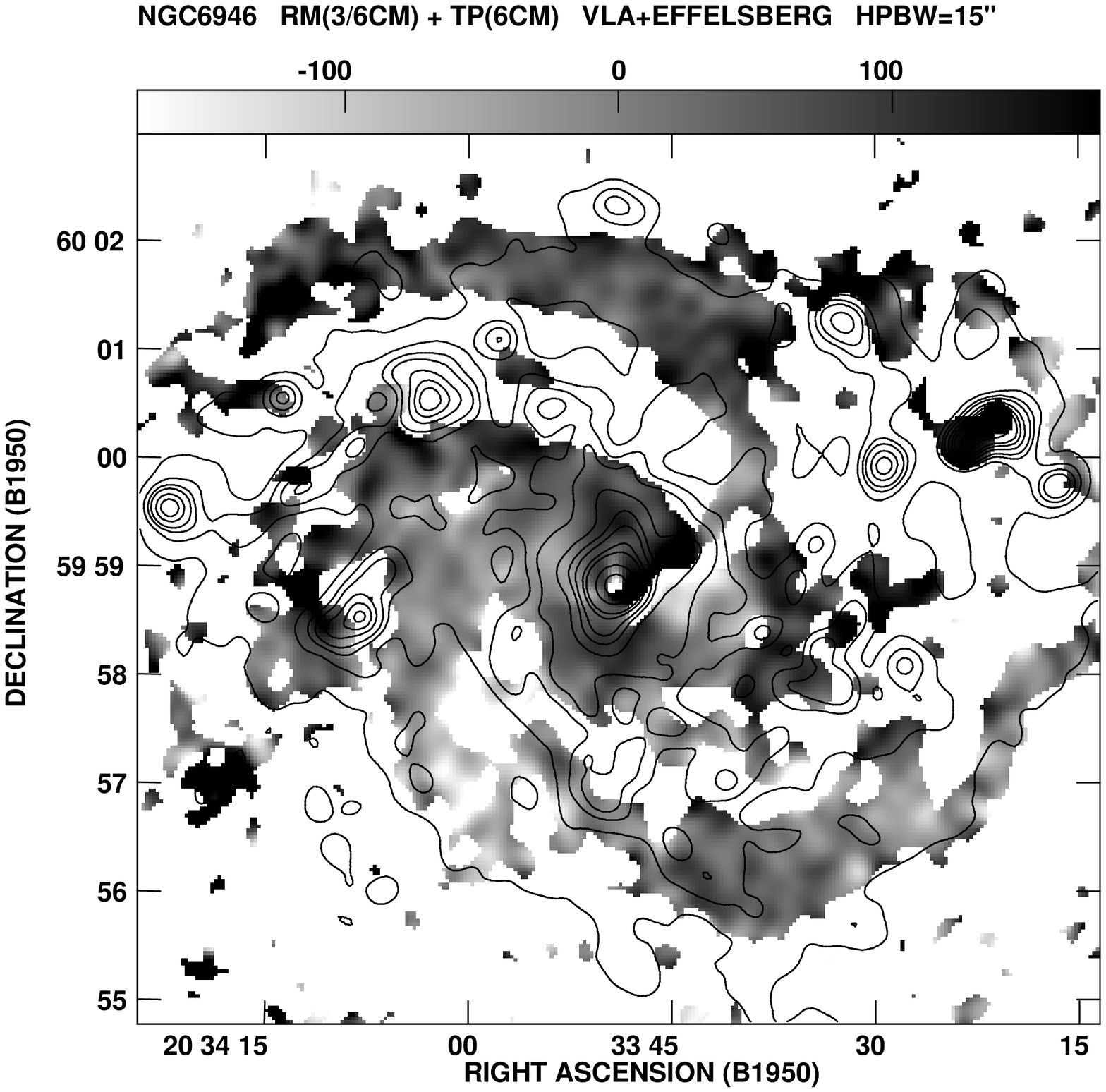,width=12.0truecm,clip=}}
\vspace{0.2cm}
\caption[]{Grey-scale map of the Faraday rotation measures of
NGC~6946 between $\lambda$3~cm and $\lambda$6~cm, overlayed onto a
total intensity map at $\lambda$6~cm. The data were obtained from
combined polarization observations with the VLA and Effelsberg
telescopes (Beck, unpublished). The angular resolution is
15~arcsec.}
\end{figure}

Figure~5 gives some idea about what an observer inside NGC~6946
may see. The large-scale $RM$ pattern is detectable only for
sufficiently long pathlengths, e.g. along the local magnetic arm
in the Galaxy or along the inner Sagittarius arm. On smaller
scales fluctuations in $RM$ with frequent reversals may dominate.
Indeed, regions with the same sign of $RM$ of several kpc in
diameter are visible in Figure~2. We need a realistic model of the
field which can be used to predict what the internal observer
would see, given a limited number of pathlengths to pulsars.
Sensitive Zeeman splitting data of a large number of clouds will
also help.

\section{Turbulent Fields}

The Galactic magnetic field cannot be completely regular because
we still observe synchrotron emission when looking at the field
along the local arm so that the regular field component in the sky
plane vanishes. Starlight and synchrotron polarization data give a
ratio of regular to turbulent field strengths of $\simeq 0.6 -
1.0$ (Section~2.1). With $B_{\mathrm reg} \simeq 4\,\mu$G the
strength of the local turbulent field is $B_{\mathrm turb} \simeq
5\,\mu$G. This agrees well with the estimates from the dispersion
of pulsar $RMs$ (\opencite{RandandKulkarni1989};
\opencite{OhnoandShibata1993}). Note that according to experience
from external galaxies the turbulent field is larger than the
regular field in the spiral arms.

Magnetic turbulence occurs over a large spectrum of scales
(\opencite{MinterandSpangler1996}). The largest scale of the
turbulent field was determined from pulsar $RMs$ as
$l_{\mathrm turb} \simeq 55$~pc (\opencite{RandandKulkarni1989}) or
$l_{\mathrm turb} \simeq 10 - 100$~pc
(\opencite{OhnoandShibata1993}). $l_{\mathrm turb}$ can also be
derived from the depolarization by turbulent fields at centimeter
radio wavelengths; the result is $l_{\mathrm turb} \simeq 25\, {\rm
pc} \times f^{1/3}$ for the galaxy NGC~6946 where $f$ is the
filling factor of the turbulent cells (\opencite{Becketal1999}).
At decimeter radio wavelengths turbulent fields also cause Faraday
dispersion. Applying standard formulae
(\opencite{Sokoloffetal1998}) yields $l_{\mathrm turb}\simeq 7\,
{\rm pc}/f$ (\opencite{Becketal1999}). The two estimates agree for
$l_{\mathrm turb}\simeq 20$~pc and $f\simeq 0.4$.

The type of turbulence (3-D or 2-D, or else) can be derived from
the {\it structure function} (\opencite{MinterandSpangler1996}).
The application to rotation measures and emission measures ($EM  =
\int  n_{\mathrm e}^2   dl$) for the same field near the Galactic
plane indicated 2-D turbulence (thin sheets or filaments), an {\it
anisotropic} turbulent field. Application to external galaxies
gave similar results (\opencite{Becketal1999}).

Future $\lambda6$~cm polarization surveys of the Galactic plane
may show the filamentary structure of the Galactic field directly.

\section{Magnetic Fields in Thick Disks and Halos}

Synchrotron emission can be followed until high Galactic latitudes
indicating the existence of a radio halo or a thick disk. Its
vertical full equivalent thickness is $3.0 \pm 0.2$~kpc near the
Sun (\opencite{Beuermannetal1985}, scaled to a distance to the
Galactic center of 8.5~kpc) which corresponds to an exponential
scale height of $h_{\mathrm syn} = 1.5 \pm 0.1$~kpc. This is also
the lower limit for the scale height of the total magnetic field.
In case of equipartition between cosmic rays and magnetic fields
and a synchrotron spectral index of $\alpha \simeq 1$, the
thickness of the distribution of the total field is $(3+\alpha)
\simeq 4$ times larger than that of the synchrotron disk, i.e.
$h_{B} \simeq 6$~kpc. The scale height of the regular field in the
warm ionized medium is $h(B_{\mathrm reg}) \simeq 1.5$~kpc,
determined from pulsar $RMs$ (\opencite{HanandQiao1994}).

In external galaxies seen edge-on the scale height of the thick
synchrotron disks is $\simeq 2$~kpc (\opencite{DumkeandKrause1998}),
similar to that of the Galaxy. The degree of polarization
increases with increasing distance from the plane, due to a decrease of
the thermal contribution and of the depolarization by turbulent fields
away from the plane. Hence the scale height of the regular field is
difficult to determine.

The local Galactic field is oriented mainly parallel to the plane,
with a vertical component of only $B_z \simeq 0.2 - 0.3~\mu$G
(\opencite{HanandQiao1994}). This agrees well with the results from
external galaxies (\opencite{Dumkeetal1995}).

Plane-parallel fields are a signature of symmetric dynamo modes
where the toroidal component has the same sign above and below the
plane and the poloidal field is of quadrupole type. $RMs$ of
extragalactic sources (see Figure~3) as well as pulsar $RMs$ have
the same sign above and below the plane for Galactic longitudes
between $90$\degr and $270$\degr. Thus the local field is part of
a large-scale symmetric (quadrupole) field structure in the thin
disk. However, towards the inner Galaxy (between $270$\degr and
$90$\degr longitude) the signs are opposite above and below the
plane. This may be due to local spurs and field loops
(\opencite{Vallee1996}), but may also indicate a global
antisymmetric (dipole) mode in the thick disk
(\opencite{AndreasyanandMakarov1988}) or in the inner galaxy
(\opencite{Hanetal1997}) where the field component perpendicular
to the plane is strong, as observed near the Galactic center
(Section~6).

Signs of an antisymmetric field were found in the huge radio halo
of the galaxy NGC~4631
(\opencite{Beck2000}). The poloidal field component is of dipole type
and allows fast propagation of cosmic rays into the halo. As a
consequence, the radio halo of NGC~4631 is larger than that of all
other edge-on galaxies observed so far.

The existence of quadrupole and/or dipole-type fields in the Galaxy
is of vital interest for cosmic-ray physics. An extension of the
$RM$ data base is required to clarify this question.

\section{Magnetic Fields Near the Galactic Center\vspace{-0.2cm}}

At $0\fdg 2$ Galactic longitude (i.e. 30~pc from the Galactic
center) a system of polarized filaments (the ``Arc'') extends to at
least $3$\degr (450~pc) {\it perpendicular} to the plane
(\opencite{Seiradakisetal1985}; \opencite{Tsuboietal1986};
\opencite{Pohletal1992}; \opencite{Haynesetal1992}). $\lambda$9~mm
Effelsberg polarization observations (Figure~6) show that the
regular magnetic field is almost perfectly aligned along the
filaments. The standard equipartition estimate cannot be applied
to derive the field strength because the electron spectrum appears
{\it monoenergetic}. Using the fact that the filaments remain
straight in the surrounding cloudy medium, the field strength was
estimated to be about 1~mG (\opencite{Morris1994}). Acceleration
and lifetime arguments led to a similar estimate
(\opencite{Reichetal2000}). This would allow confinement of
high-energy cosmic rays, but requires an effective acceleration
process, e.g. via reconnection (\opencite{LeschandReich1992}). A
monoenergetic electron spectrum can also be provided by shock
acceleration if the particle escape time is much longer than the
acceleration time (\opencite{Schlickeiser1984}).

\begin{figure}
\vspace{0.2cm}
\centerline{\psfig{file=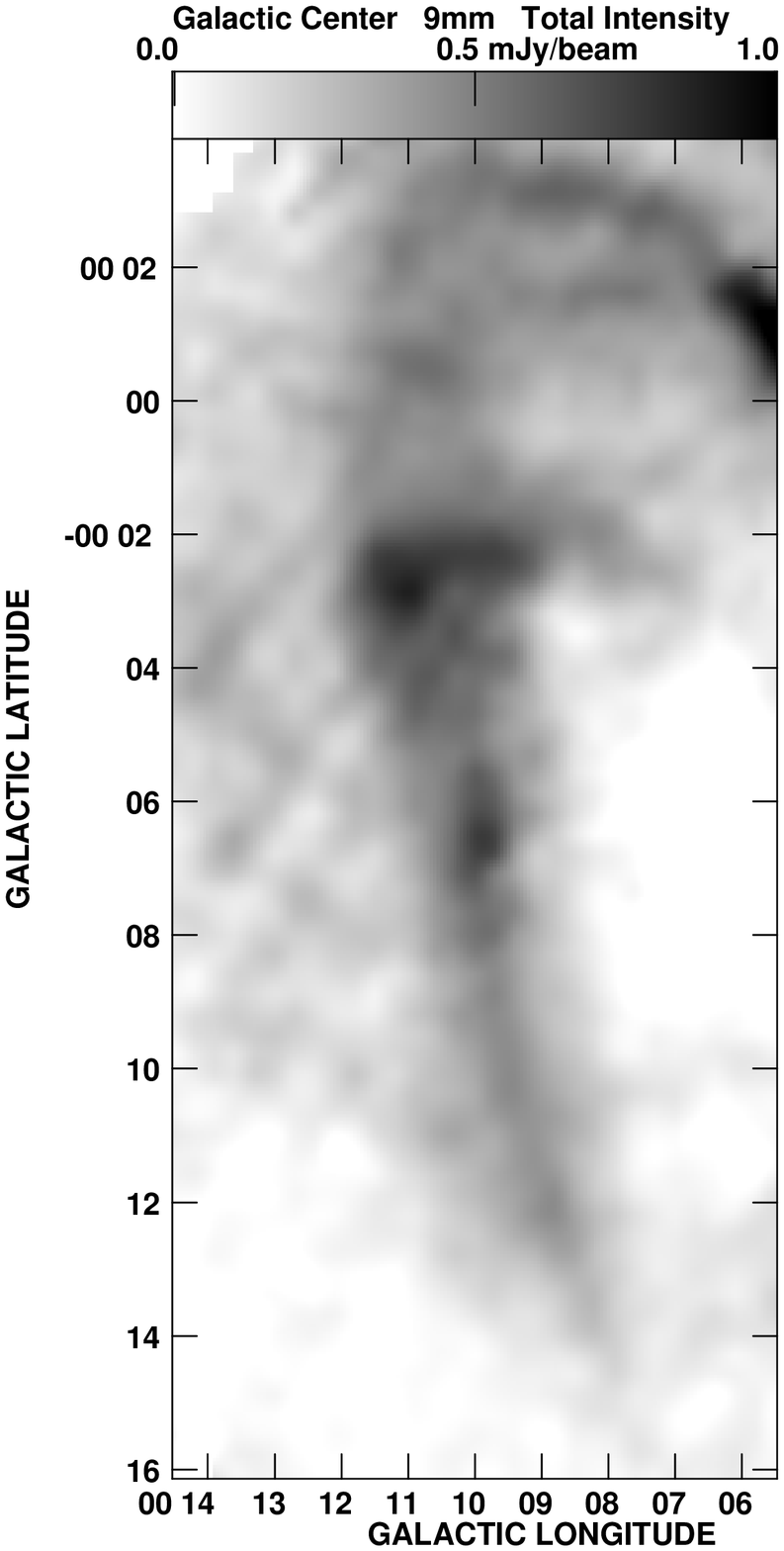,width=6.0truecm,clip=}
\hfill
\psfig{file=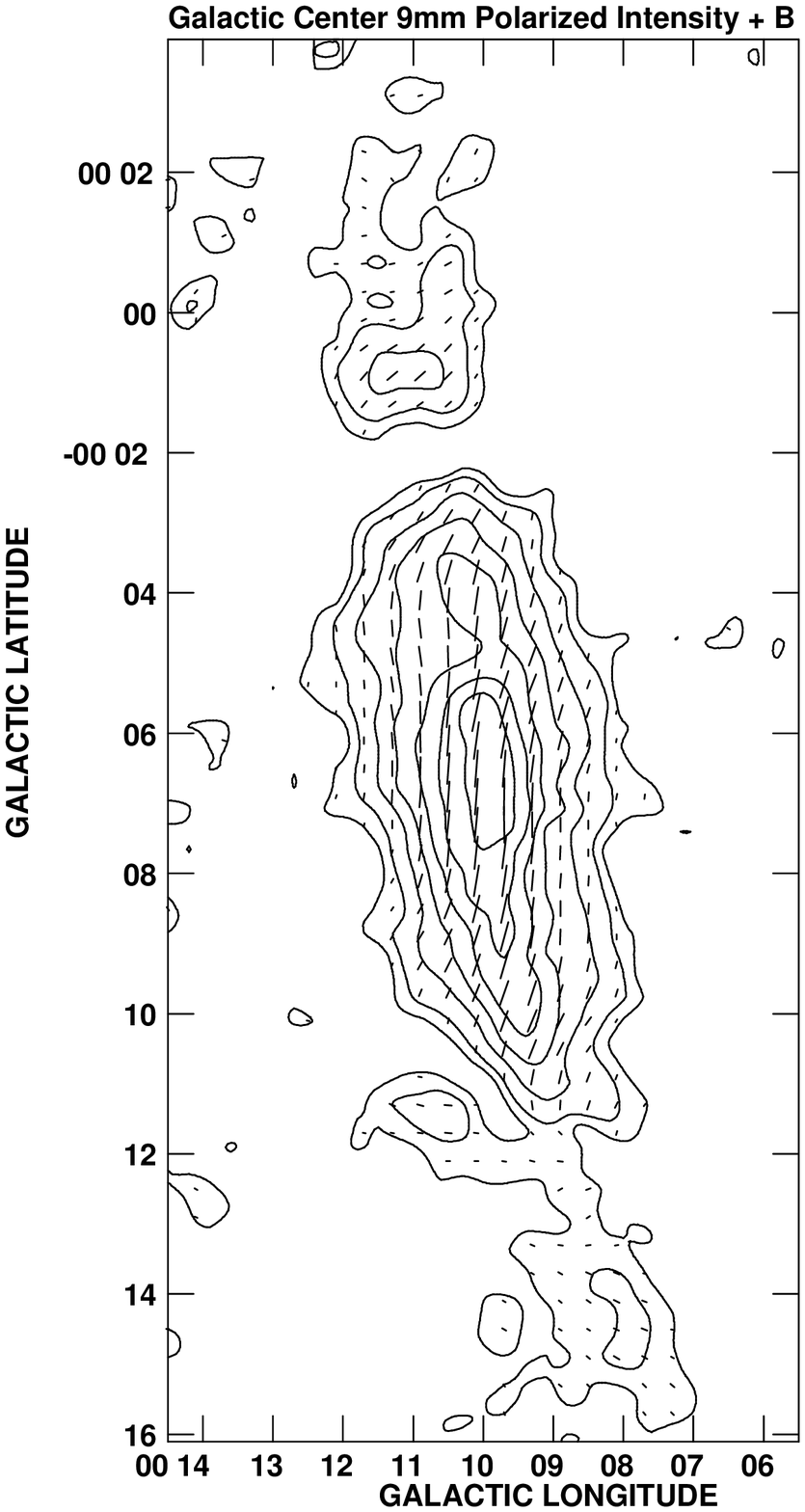,width=6.0truecm,clip=}}
\vspace{0.2cm}
\caption[]{Total radio emission (left) and polarized emission
with $\mathbf{B}$--vectors (right) from the inner part of
the ``Arc'' located $0\fdg 2$ east from the
Galactic center, observed at $\lambda$9~mm with the Effelsberg
telescope (from \opencite{LeschandReich1992}). The angular
resolution is 36~arcsec.}
\end{figure}

Also in a few dense gas clouds in front of the Galactic center (about 2~pc
north of the Galactic center) field strengths in the milligauss range
were derived from Zeeman measurements (\opencite{Planteetal1994};
\opencite{Yusef-Zadehetal1996}). On the other hand, low
observed bandwidth depolarization restricts the {\it average} field in the
SgrA complex to less than 0.4~mG (\opencite{Reich1994}). The non-detection
of the Zeeman effect in the OH lines (\opencite{UchidaandGuesten1995})
also indicates a relatively weak general magnetic field into which
bundles or clouds with strong fields are embedded.

Vertical fields in the central regions were detected in several spiral
galaxies (\opencite{Hummeletal1983}), but were mostly assigned
to nuclear activity. Radio polarization observations of edge-on
galaxies with high resolution may show similarities to
the central region of the Galaxy.

Recently a filament {\it parallel} to the Galactic plane,
comprised of parallel strands, has been found $1\fdg 5$ from the
center (\opencite{Langetal1999}), possibly linking the central
with the global field (Section~5).

\section{Summary}

\begin{itemize}
\item The strength of the total magnetic field in the Galaxy is
$6\pm2\,\mu$G locally and $10\pm3\,\mu$G at 3~kpc Galactic
radius. The local equipartition value agrees well with that derived
using direct measurements of cosmic rays and $\gamma$-rays.
\item The strength
of the local regular field is $4\pm1\mu$G, based on optical and synchrotron
polarization data, while pulsar rotation measures give a more than twice
lower value.  The optical/synchrotron data may overestimate the regular
field strength due to the presence of anisotropic fields, while the pulsar
value may be an underestimate due to anticorrelated
fluctuations in regular field strength and in thermal electron density.
\item The local regular field may be part of a ``magnetic arm'' between
the optical arms. The global structure of the Galactic field is unknown.
\item No large-scale field reversals similar to those observed in the
Galaxy were found in external galaxies. Some of the
Galactic reversals may be due to large-scale anisotropic field loops.
More pulsar $RM$ and Zeeman splitting data are needed.
\item Polarization surveys of the Galactic plane at decimeter wavelengths
show a wealth of details which are mostly due to variations of Faraday
rotation measures in the foreground medium. In order to directly observe
the fine structure of the field, surveys at wavelengths $\le$6~cm are
required.
\item The Galaxy is surrounded by a thick radio disk with a scale height
of $\simeq 1.5$~kpc, similar to that of edge-on spiral galaxies.
\item The local Galactic field in the thin disk is of even symmetry with
respect to the plane (quadrupole), but a global dipole field in the
thick disk is possible.
\item The Galactic center region hosts highly regular fields of up to
milligauss
strength which are concentrated in narrow bundles, oriented perpendicular
to the plane.
\end{itemize}

\begin{acknowledgements}
The author thanks Dr. E.\,M.\,Berkhuijsen for providing Figure~1
and for careful reading of the manuscript. Drs. JinLin Han,
Wolfgang Reich, John Seiradakis, Steve Spangler, Andy Strong and
Richard Wielebinski are acknowledged for many useful comments.
\end{acknowledgements}

{}
\end{article}
\end{document}